\documentclass[]{eptcs}
\providecommand{\event}{QPL 2014} 
\usepackage{breakurl}             

\usepackage{amsmath,amssymb,amsthm}
\usepackage{graphicx}
\usepackage{cite}
\usepackage{braket}
\usepackage{multicol}
\newtheorem{lemm}{Lemma}
\newtheorem{prop}{Proposition}

\theoremstyle{definition}
\newtheorem{defi}{Definition}
\theoremstyle{remark}
\newtheorem{exam}{Example}

\newcommand{\trans}[1]{\overset{#1}{\longrightarrow}}
\newcommand{\Trans}[1]{\overset{#1}{\Longrightarrow}}
\newcommand{\con}[2]{\left<#1, #2\right>}
\newcommand{\ifthen}[2]{\mathbf{if}\ #1\ \mathbf{then}\ #2}

\title{Observational Equivalence Using Schedulers for Quantum Processes}
\author{
Kazuya Yasuda \qquad\qquad Takahiro Kubota \quad\quad Yoshihiko Kakutani
\institute{Dept. of Computer Science\\
Graduate School of Information Science and Technology\\ The University of Tokyo\\
Tokyo, Japan}
\email{\{kyasuda, takahiro.k11\_30, kakutani\}@is.s.u-tokyo.ac.jp}}
\def\titlerunning{Observational Equivalence Using Schedulers for Quantum Processes}
\def\authorrunning{K.Yasuda, T. Kubota \& Y. Kakutani}

\begin{document}

\maketitle


\begin{abstract}

In the study of quantum process algebras, researchers have introduced different
notions of equivalence between quantum processes like bisimulation or barbed
congruence. However, there are intuitively equivalent quantum processes that
these notions do not regard as equivalent.
In this paper, we introduce a notion of equivalence named observational
equivalence into qCCS. Since quantum processes have both probabilistic and
nondeterministic transitions, we introduce schedulers that solve nondeterministic
choices and obtain probability distribution of quantum processes.
By definition, restrictions of schedulers change observational equivalence.
We propose some definitions of schedulers, and investigate the relation between
the restrictions of schedulers and observational equivalence.

\end{abstract}


\section{Introduction}

Quantum communication protocols have been proposed since Bennett and Brassard
\cite{Bennett1984} proposed a quantum key distribution (QKD) protocol.
However, proving the correctness or security of communication protocols is very
complicated and error-prone because quantum mechanical behavior is often
different from our intuition based on classical mechanics. In order to analyze
or verify quantum protocols successfully, quantum process calculi have been
proposed, for example, QPAlg \cite{Jorrand2004}, CQP \cite{Gay2005}, and
qCCS \cite{Feng2007,Feng2012,Ying2009}.

In quantum process calculi, it is one of the important notions whether two
processes behave similarly or not, in other words, whether they are behaviorally
equivalent or not. One of the benefits of this notion is to provide the
following technique to verify the correctness of a communication protocol.
First, write a process that models the procedure of the communication protocol.
Second, define a simpler process that is the specification of the protocol.
Then, if these two processes are behaviorally equivalent, it is proved that the
protocol satisfies the specification. For instance, the correctness of
quantum teleportation is shown by using qCCS in \cite{Feng2012}. In the
paper, the model and the specification of quantum teleportation are defined
with qCCS as $Tel = (Alice_t || Bob_t) \backslash \{e\}$ and
$Tel_{spec} = \mathtt{c}?q.
SWAP_{1,3}[q, q_1, q_2].\mathtt{d}!q_2.\mathbf{nil}$ respectively, where
\[
 Alice_t := \mathtt{c}?q.CN[q, q_1].H[q].
 M[q, q_1; x].e!x.\mathbf{nil},\ 
 Bob_t := e?x.\sum_{0 \le i \le 3}\left(\ifthen{x = i}
 {\sigma^i[q_2].\mathtt{d}!q_2.\mathbf{nil}}\right),
\]
and it is proved that the model $Tel$ and the specification $Tel_{spec}$
are behaviorally equivalent (definitions of some symbols are in
\cite{Feng2012}).

There is a variety of the notions of behavioral equivalence such as
(weak) bisimulation and barbed congruence. For example, these notions for
qCCS are defined in \cite{Deng2012,Feng2012}. Intuitively, bisimulation
is the notion that one process can simulate the other's behavior, and
barbed congruence is the notion that any observers (or attackers) cannot
distinguish two processes.

These two notions have widely been used in formal verification of
processes. However, there are some processes that are not regarded as
equivalent by these notions but intuitively equivalent. This problem
occurs when the processes include quantum operations or
communication. For example, consider the following two processes: one
sends a qubit $\ket{0}$ or $\ket{1}$ with the same probability, the
other sends $\ket{+}$ or $\ket{-}$ with the same probability. These two
processes are not regarded as equivalent by the notion of
bisimulation. However, we have intuitively regarded these two processes
as the same process because these qubits are expressed as the same
density matrix. This kind of equation was used in the security proof of
BB84 by Shor and Preskill \cite{Shor2000}.

The aim of this paper is to define the notion of equivalence that regards above
cases as equivalent into the quantum process calculus qCCS. This notion is called
observational equivalence. Intuitively, two processes are observationally
equivalent when they are observed the same by any attackers. Because attackers
can observe their behavior only by watching the channels that they use, processes are
observed the same when they use the channels with the same probability. In
addition, we must consider the probability of using channels although the
quantum processes of qCCS have both probabilistic and nondeterministic
transitions. In order to solve this inconvenience, we define schedulers that
solve nondeterministic choices and obtain probability distribution of
quantum processes. By definition, the restrictions of schedulers change
observational equivalence. We propose some definitions of schedulers, and
investigate the relation between the restrictions of schedulers and
observational equivalence.

\section{Definitions of qCCS}

In this section, we introduce the language qCCS proposed in 
\cite{Feng2007,Feng2012,Ying2009}.

\subsection{Syntax}

Three types of data are considered in qCCS: \texttt{Bool} for booleans,
\texttt{Real} for real numbers and \texttt{Qbt} for qubits.
Let $cVar$ be the set of classical variables, ranged over by $x, y, \dots$,
and $qVar$ be the set of quantum variables, ranged over by $q, r, \dots$.
We assume that $cVar$ and $qVar$ are both countably infinite and 
$cVar \cap qVar = \emptyset$.
The indexed set $\{q_1, \dots, q_n\}$ is often abbreviated to $\tilde{q}$.
Let $Exp$ be the set of classical data expressions over \texttt{Real}, 
ranged over by $e, e', \dots$, which includes $cVar$ as a subset.
Let $BExp$ be the set of boolean-valued expressions, ranged over by $b, b', \dots$. 

Two types of channels are used in qCCS: $cChan$ for classical channels and 
$qChan$ for quantum channels. $c, d, \dots$ range over $cChan$ and 
$\mathtt{c}, \mathtt{d}, \dots$ range over $qChan$. 
We assume that $cChan \cap qChan = \emptyset$.
Let $Chan$ be the set of all channels, that is, $Chan = cChan \cup qChan$.
A \textit{relabeling function} is a function $f: Chan \to Chan$ such that 
$f(cChan) \subset cChan$ and $f(qChan) \subset qChan$.

The set of \textit{quantum processes} $qProc$ is defined inductively as follows:
\begin{eqnarray*}
qProc \ni P,Q &::=& \mathbf{nil}\ |\ A(\tilde{q};\tilde{x})\ |\ \tau.P\ |\ c?x.P\ |\ 
c!e.P\ |\ \mathtt{c}?q.P\ |\ \mathtt{c}!q.P\ |\ \mathcal{E}[\tilde{q}].P\ |\ 
M[\tilde{q};x].P\ |\ \\
&& P+Q\ |\ P||Q\ |\ P[f]\ |\ P \backslash L\ |\ 
\mathbf{if}\ b\ \mathbf{then}\ P
\end{eqnarray*}
where $c \in cChan$, $x \in cVar$, $e \in Exp$, 
$\mathtt{c} \in qChan$, $b \in BExp$, $q \in qVar$, 
$A(\tilde{q};\tilde{x})$ is a process constant, $\tau$ is the silent action,
$f$ is a relabeling function, $L \subset_\mathrm{fin} Chan$, 
$\mathcal{E}$ and $M$ are respectively a trace-preserving super-operator and 
a non-degenerate projective measurement applying on the Hilbert space 
associated with the systems $\tilde{q}$. 
The process $\mathbf{nil}$ may be omitted, for instance, $c!0$ is used instead of 
$c!0.\mathbf{nil}$.

The \textit{free classical variable function} $fv: qProc \to 2^{cVar}$ is defined 
in the usual way. Note that the quantum measurement $M[\tilde{q};x]$ binds
the variable $x$, that is, $fv(M[\tilde{q};x].P) = fv(P) - \{x\}$.
A process $P$ is \textit{closed} if $fv(P) = \emptyset$.
The \textit{free quantum variable function} $qv: qProc \to 2^{qVar}$ is defined 
inductively as in Figure \ref{fig:qv}.
\begin{figure}[tb]
\begin{multicols}{3}
\begin{center}
{\small
$qv(\mathbf{nil}) = \emptyset$ \\
$qv(A(\tilde{q};\tilde{x})) = \tilde{q}$ \\
$qv(\tau.P) = qv(P)$ \\
$qv(c?x.P) = qv(P)$ \\
$qv(c!e.P) = qv(P)$ \\
$qv(\mathtt{c}?q.P) = qv(P) - \{q\}$ \\
$qv(\mathtt{c}!q.P) = qv(P) \cup \{q\}$ \\
$qv(\mathcal{E}[\tilde{q}].P) = qv(P) \cup \tilde{q}$ \\
$qv(M[\tilde{q};x].P) = qv(P) \cup \tilde{q}$ \\
$qv(P+Q) = qv(P) \cup qv(Q)$ \\
$qv(P||Q) = qv(P) \cup qv(Q)$ \\
$qv(P[f]) = qv(P)$ \\
$qv(P \backslash L) = qv(P)$ \\
$qv(\ifthen{b}{P}) = qv(P)$
}
\end{center}
\end{multicols}
\caption{Definition of $qv$}
\label{fig:qv}
\end{figure}
For quantum processes to be legal, we require that 
\begin{enumerate}
\item
$q \not\in qv(P)$ in the process $\mathtt{c}!q.P$;
\item
$qv(P) \cap qv(Q) = \emptyset$ in the process $P || Q$;
\item
each process constant $A(\tilde{q};\tilde{x})$ has a defining equation 
$A(\tilde{q};\tilde{x}) := P$, where $P \in qProc$, $qv(P) \subset \tilde{q}$ 
and $fv(P) \subset \tilde{x}$.
\end{enumerate}
We use $P\{v/x\}$ to denote the substitution of $v$ for $x$ in $P$. 
We abbreviate $P\{v_1/x_1\}\dots\{v_n/x_n\}$ to $P\{\tilde{v}/\tilde{x}\}$.

\subsection{Configuration}

For each $q \in qVar$, we assume a 2-dimensional Hilbert space $\mathcal{H}_q$ 
to be the state space associated with the system $q$. 
Let 
$$\mathcal{H}_S = \bigotimes_{q \in S} \mathcal{H}_q$$
for any $S \subset qVar$.
In particular, $\mathcal{H} = \mathcal{H}_{qVar}$ is the whole state space 
associated with all of the quantum variables.

A \textit{configuration} is a pair $\con{P}{\rho}$, where $P \in qProc$ is closed and 
$\rho$ is a density operator on $\mathcal{H}$. 
Let $Con$ be the set of all configurations, 
ranged over by $C, D, \dots$.
If the state associated with the system $q$ is $\ket{\psi}\bra{\psi}$ , 
the notation $\ket{\psi}\bra{\psi}_q \otimes \rho$ or $[\ket{\psi}]_q \otimes \rho$
is used to denote this whole state, where $\rho$ is a state associated with the systems 
$qVar - \{q\}$.

Let $D(Con)$ be the set of finite-support probability distribution over $Con$, 
ranged over by $\mu, \nu, \dots$.
When $\mu(C) = 1$ for some $C \in Con$, we use $C$ instead of $\mu$ to
denote the distribution. We sometimes use a form
$\mu = \boxplus_{i \in I} p_i \bullet C_i$ to denote the distribution
$\mu$, where $C_i$ are distinct elements of $Con$ and $\mu(C_i) = p_i$.
For any $\mu = \boxplus_{i \in I} p_i \bullet \con{P_i}{\rho_i}$ and
trace-preserving super-operator $\mathcal{E}$, the notation 
$\boxplus_{i \in I} p_i \bullet \con{P_i}{\mathcal{E}(\rho_i)}$ 
is often abbreviated to $\mathcal{E}(\mu)$.

\subsection{Operational semantics}

Let 
$Act = \{\tau\} \cup \{c?v, c!v\ |\ c \in cChan, v \in \mathtt{Real}\} \cup 
\{\mathtt{c}?r, \mathtt{c}!r\ |\ \mathtt{c} \in qChan, r \in qVar\}$.
For each $\alpha \in Act$, let $cn(\alpha)$ be the set of channel names used in 
the action $\alpha$, that is, 
$cn(\tau) = \emptyset$, $cn(c?v) = cn(c!v) = \{c\}$ and 
$cn(\mathtt{c}?r) = cn(\mathtt{c}!r) = \{\mathtt{c}\}$. 
For each $\alpha \in Act$ and relabeling function $f$, we use $f(\alpha)$ to denote 
the action of which channel is relabeled by $f$. For example, 
$f(\tau) = \tau$, $f(c?v) = f(c)?v$ and $f(\mathtt{c}!q) = f(\mathtt{c})!q$.

The operational semantics of qCCS is defined by the probabilistic labeled 
transition system \cite{Deng2012} $(Con, Act, \longrightarrow)$, 
where $\longrightarrow\ \subset Con \times Act \times D(Con)$ is the smallest relation 
satisfying the rules defined in Figure \ref{fig:transrule} (the symmetric forms for 
rules \textsc{C-Com}, \textsc{Q-Com}, \textsc{Inp-Int}, \textsc{Oth-Int} 
and \textsc{Sum} are omitted). 
Here, $[\![e]\!]$ and $[\![b]\!]$ are the usual interpretations of $e \in Exp$ and 
$b \in BExp$ respectively, and $\mathcal{E}_{\tilde{q}}$ means that the super-operator 
$\mathcal{E}$ applys on the state associated with the systems $\tilde{q}$.
We write $C \trans{\alpha} \mu$ instead of $(C, \alpha, \mu) \in 
\longrightarrow$. 
We write $C \trans{\alpha}$ when there exists $\mu \in D(Con)$ such that 
$C \trans{\alpha} \mu$. 
We write $C \not\rightarrow$ when there do not exist $\alpha$ and $\mu$ such 
that $C \trans{\alpha} \mu$.

\begin{figure}[tb]
\begin{center}
\includegraphics[width=135mm,bb=70 400 540 720,clip]{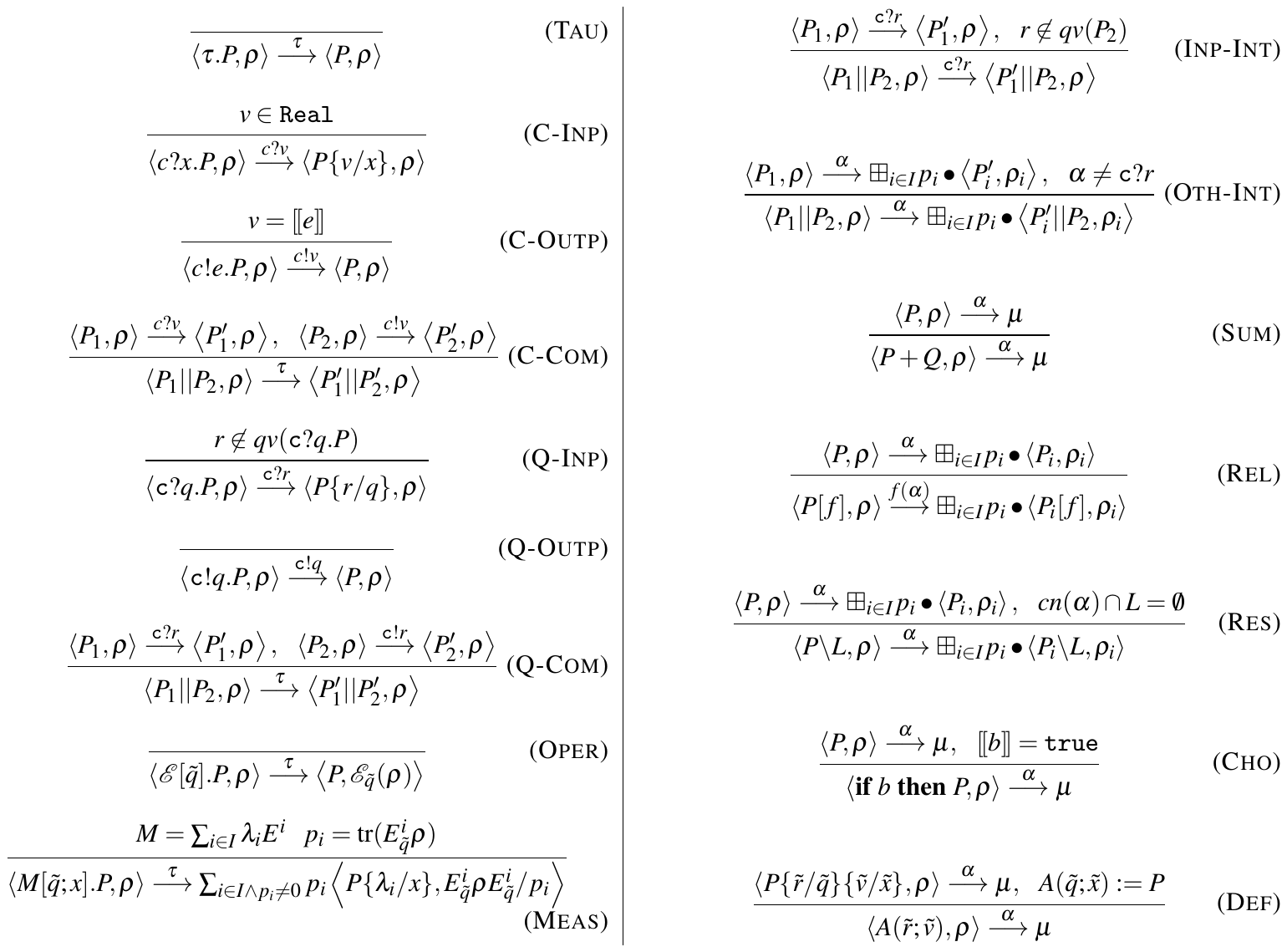}
\end{center}
\caption{Transition rules of qCCS}
\label{fig:transrule}
\end{figure}

The transition relation $\longrightarrow$ is lifted to 
$D(Con) \times Act \times D(Con)$ as follows: we write $\mu \trans{\alpha} \nu$ if 
for any $C \in supp(\mu)$, $C \trans{\alpha} \nu_C$ for some 
$\nu_C$, and 
$\nu = \sum_{C \in supp(\mu)} \mu(C)\nu_C$.

\section{Bisimulation}

In this section, we recall the relation called open bisimulation.
To define it, we need to define the relation $\Longrightarrow$ and 
a weight function. These definitions are introduced in
\cite{Feng2012}.

\begin{defi}
 The relation $\Longrightarrow\ \subset D(Con) \times D(Con)$ is the
 smallest relation satisfying the following conditions:
 \begin{enumerate}
  \item $C \Longrightarrow C$;
  \item if $C \trans{\tau} \mu$ and $\mu \Longrightarrow \nu$, then $C
	\Longrightarrow \nu$;
  \item if $\mu = \sum_{i \in I} p_i C_i$, and for any $i \in I$,
	$C_i \Longrightarrow \nu_i$ for some $\nu_i$, then
	$\mu \Longrightarrow \sum_{i \in I} p_i \nu_i$.
 \end{enumerate}

 For any $\mu, \nu \in D(Con)$ and
 $s = \alpha_1\dots\alpha_n \in Act^*$, we say that $\mu$ can evolve
 into $\nu$ by a weak $s$-transition, denoted by $\mu \Trans{s} \nu$, if
 there exist $\mu_1, \dots, \mu_{n+1}, \nu_1, \dots, \nu_n \in D(Con)$,
 such that $\mu \Longrightarrow \mu_1$, $\mu_{n+1} = \nu$, and
 for each $i = 1, \dots, n$, $\mu_i \trans{\alpha_i} \nu_i$ and
 $\nu_i \Longrightarrow \mu_{i+1}$.

 For any $s \in Act^*$, $\hat{s}$ is the string obtained from $s$ by
 deleting all the occurrences of $\tau$.
\end{defi}

\begin{defi}
Let $\mathcal{R} \subset Con \times Con$ and $\mu, \nu \in D(Con)$. 
A \textit{weight function} for $(\mu, \nu)$ w.r.t. $\mathcal{R}$ is a function 
$\delta: Con \times Con \to [0, 1]$ that satisfies the following conditions: 
\begin{enumerate}
\item
for all $C, D \in Con$, 
\[
\sum_{D' \in supp(\nu)} \delta(C,D') = \mu(C),
\sum_{C' \in supp(\mu)} \delta(C',D) = \nu(D);
\]
\item
for all $C, D \in Con$, 
if $\delta(C,D) > 0$, then $C \mathcal{R} D$.
\end{enumerate}
We write $\mu\mathcal{R}\nu$ if there exists a weight function for $(\mu,\nu)$ w.r.t. 
$\mathcal{R}$.
\end{defi}

\begin{lemm}
Let $\mu, \nu \in D(Con)$. Then $\mu \mathcal{R} \nu$ if and only if 
there exist $\{p_i\}_{i \in I}$, $\{C_i\}_{i \in I}$, and 
$\{D_i\}_{i \in I}$ such that  
$\mu = \sum_{i \in I} p_iC_i$, $\nu = \sum_{i \in I} p_iD_i$, 
and $C_i\mathcal{R}D_i$ for each $i \in I$. 
In particular, if $C\mathcal{R}\mu$ then $C\mathcal{R}D$ 
for each $D \in supp(\mu)$.
\end{lemm}

Now we introduce open bisimulation on qCCS defined in \cite{Deng2012}.

\begin{defi}
A relation $\mathcal{R} \subset Con \times Con$ is an \textit{open bisimulation} 
if $\con{P}{\rho} \mathcal{R} \con{Q}{\sigma}$ implies that $qv(P) = qv(Q)$,
$\mathrm{tr}_{qv(P)}(\rho) = \mathrm{tr}_{qv(Q)}(\sigma)$, and 
for any super-operator $\mathcal{E}$ acting on $\mathcal{H}_{\overline{qv(P)}}$,
\begin{enumerate}
\item
whenever $\con{P}{\mathcal{E}(\rho)} \trans{\alpha} \mu$, there exists $\nu$ such that 
$\con{Q}{\mathcal{E}(\sigma)} \Trans{\hat{\alpha}} \nu$ and $\mu\mathcal{R}\nu$;
\item
whenever $\con{Q}{\mathcal{E}(\sigma)} \trans{\alpha} \nu$, there exists $\mu$ such that 
$\con{P}{\mathcal{E}(\rho)} \Trans{\hat{\alpha}} \mu$ and $\mu\mathcal{R}\nu$.
\end{enumerate}

Let $\approx_o$ be the largest open bisimulation.
\end{defi}

There are other notions of equivalence like open bisimulation on qCCS. 
For example, \textit{bisimulation} is defined in 
\cite{Feng2012} and 
\textit{reduction barbed congruence} is defined in \cite{Deng2012}. 
According to \cite{Deng2012}, the largest open bisimulation is strictly coarser than 
the largest bisimulation, and the reduction barbed congruence coincides with the largest 
open bisimulation.

\section{Observational equivalence}

In this section, we introduce the notion of observational equivalence on qCCS. 
Intuitively, two configurations are observationally equivalent 
when they are observed by foreign processes in the same way, in other words, 
when they use the same channels with the same probability in any contexts.

First of all, we describe why we want to define the notion of observational equivalence 
with an example.
There are two different ways to express quantum measurements in qCCS: 
$M[q;x]$ and $\mathcal{E}[q]$, where $M$ is the 1-qubit projective measurement such that 
$M = \sum_{i=0}^1 i \ket{i}\bra{i}$, $\mathcal{E}$ is the trace-preserving super-operator 
such that $\mathcal{E}(\rho) = \sum_{i=0}^1 \ket{i}\bra{i} \rho \ket{i}\bra{i}$. 
We intuitively want to consider that these two processes are equivalent, but they are not 
bisimilar. This gap is an obstacle to formalize Shor and Preskill's security proof of 
BB84 \cite{Kubota2012}. 
For simplicity, we consider the following example.

\begin{exam}\label{examC}
Consider these two configurations:
\[
C = \con{M[q;x].(c!0+d!0)}{[\ket{+}]_q \otimes \rho}, 
\hspace{5mm}
D = 
\con{\mathcal{E}[q].(c!0+d!0)}{[\ket{+}]_q \otimes \rho}
\]
where $M$ and $\mathcal{E}$ are described above. 
The pLTSs for these configurations are depicted as in Figure \ref{fig:plts-exampleC}.
It is obvious that $C \not\approx_o D$. We want to consider that 
$C$ and $D$ are equivalent.

\begin{figure}[tb]
 \centering
 \scalebox{0.8}{\begin{picture}(310,262)(0,0)
  \put(0,154){
   \put(0,96){\makebox(310,12){%
    $C = \con{M[q;x].(c!0+d!0)}{[\ket{+}]_q \otimes \rho}$%
   }}
   \put(155,93){\line(0,-1){20}}
   \put(160,73){\makebox(30,20)[l]{$\tau$}}
   \put(155,73){\vector(-3,-1){30}}
   \put(107,68){\makebox(30,10)[r]{\footnotesize $1/2$}}
   \put(155,73){\vector(3,-1){30}}
   \put(173,68){\makebox(30,10)[l]{\footnotesize $1/2$}}
   \put(0,48){\makebox(150,12){$\con{c!0 + d!0}{[\ket{0}]_q \otimes \rho}$}}
   \put(160,48){\makebox(150,12){$\con{c!0 + d!0}{[\ket{1}]_q \otimes \rho}$}}
   \put(35,45){\vector(0,-1){30}}
   \put(40,15){\makebox(30,30)[l]{$c!0$}}
   \put(115,45){\vector(0,-1){30}}
   \put(120,15){\makebox(30,30)[l]{$d!0$}} 
   \put(195,45){\vector(0,-1){30}}
   \put(200,15){\makebox(30,30)[l]{$c!0$}}
   \put(275,45){\vector(0,-1){30}}
   \put(280,15){\makebox(30,30)[l]{$d!0$}} 
   \put(0,0){\makebox(70,12){$\con{\mathbf{nil}}{[\ket{0}]_q \otimes \rho}$}}
   \put(80,0){\makebox(70,12){$\con{\mathbf{nil}}{[\ket{0}]_q \otimes \rho}$}}
   \put(160,0){\makebox(70,12){$\con{\mathbf{nil}}{[\ket{1}]_q \otimes \rho}$}}
   \put(240,0){\makebox(70,12){$\con{\mathbf{nil}}{[\ket{1}]_q \otimes \rho}$}}
  }
  \put(10,0){
   \put(0,112){\makebox(290,20){
    $D = \con{\mathcal{E}[q].(c!0+d!0)}{[\ket{+}]_q \otimes \rho}$
   }}
   \put(145,109){\vector(0,-1){30}}
   \put(150,79){\makebox(30,30)[l]{$\tau$}}
   \put(0,56){\makebox(290,20){
    $\con{c!0+d!0}
     {\left(\frac{1}{2}\ket{0}\bra{0}+\frac{1}{2}\ket{1}\bra{1}\right)_q\otimes\rho}$
   }}
   \put(80,53){\vector(0,-1){30}}
   \put(85,23){\makebox(30,30)[l]{$c!0$}}
   \put(210,53){\vector(0,-1){30}}
   \put(215,23){\makebox(30,30)[l]{$d!0$}}
   \put(0,0){\makebox(140,20){
    $\con{\mathbf{nil}}
     {\left(\frac{1}{2}\ket{0}\bra{0}+\frac{1}{2}\ket{1}\bra{1}\right)_q\otimes\rho}$
   }}
   \put(150,0){\makebox(140,20){
    $\con{\mathbf{nil}}
     {\left(\frac{1}{2}\ket{0}\bra{0}+\frac{1}{2}\ket{1}\bra{1}\right)_q\otimes\rho}$
   }}
  }
 \end{picture}}
 \caption{pLTSs for Example \ref{examC}}
 \label{fig:plts-exampleC}
\end{figure}
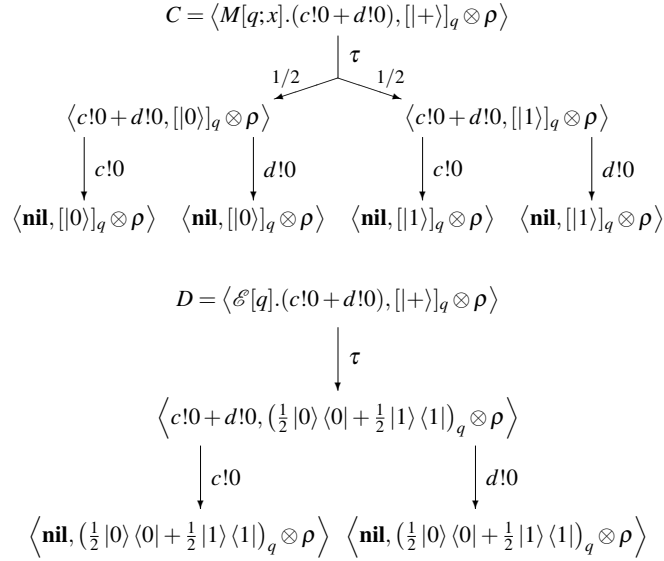

\end{exam}

\subsection{Scheduler}

Even though quantum processes on qCCS have both probabilistic and 
nondeterministic transitions, we have to consider a probability to use channels
in order to define observational equivalence.
So, we define schedulers to solve nondeterministic choices and to obtain 
probability distribution of configurations.

\begin{defi}
A function $F: Con \to (Act \times D(Con)) \cup \{\bot\}$ is a \textit{scheduler} 
if the following conditions are satisfied:
\begin{enumerate}
\item
$F(C) = (\alpha, \mu)$ implies $C \trans{\alpha} \mu$,
\item
$F(C) = \bot$ implies $C \not\rightarrow$.
\end{enumerate}

We write $C \trans{\alpha}_F \mu$ when $F(C) = (\alpha, \mu)$. 
We write $C \trans{\alpha}_F$ when $F(C) = (\alpha, \mu)$ 
for some $\mu \in D(Con)$.
\end{defi}

The relation $\Longrightarrow$ is limited by a scheduler as follows:
\begin{defi}
 The relation $\Longrightarrow_F\ \subset D(Con) \times D(Con)$ is the
 smallest relation satisfying the following conditions:
 \begin{enumerate}
  \item $C \Longrightarrow_F C$;
  \item if $C \trans{\tau}_F \mu$ and $\mu \Longrightarrow_F \nu$,
	then $C \Longrightarrow_F \nu$;
  \item if $\mu = \sum_{i \in I} p_i C_i$, and for any $i \in I$,
	$C_i \Longrightarrow_F \nu_i$ for some $\nu_i$, then
	$\mu \Longrightarrow_F \sum_{i \in I} p_i \nu_i$.
 \end{enumerate}
\end{defi}

\subsection{Observational equivalence}

We write $C \Downarrow_F^p c$ when there exists $\mu \in D(Con)$ such that 
\begin{itemize}
\item
$C \Longrightarrow_F \mu$ holds; 
\item
for each $C' \in supp(\mu)$, either $F(C') = \bot$ or 
$C' \trans{\lambda}_F$ holds for some $\lambda \not= \tau$; and 
\item
the equation
$\sum \{ \mu(C')\ |\ C' \trans{c!v}_F \mbox{ for some } v \} = p$
holds.
\end{itemize}
This means, intuitively, that the configuration $C$ uses the channel $c$ 
with the probability $p$ after all internal transitions in accordance with the 
scheduler $F$.

Now, we define observational equivalence on qCCS. 

\begin{defi}
Two configurations $\con{P}{\rho}, \con{Q}{\sigma} \in Con$ are 
\textit{observationally equivalent}, 
we write $\con{P}{\rho} \approx_{oe} \con{Q}{\sigma}$, 
if $qv(P) = qv(Q)$, $\mathrm{tr}_{qv(P)}(\rho) = \mathrm{tr}_{qv(Q)}(\sigma)$ and 
for any quantum processes $R \in qProc$,
\begin{enumerate}
 \item for each scheduler $F$ there exists a scheduler $F'$ such that,
       for any classical channel $c \in cChan$
       $\con{P || R}{\rho} \Downarrow^p_F c$ implies that
       $\con{Q || R}{\sigma} \Downarrow^p_{F'} c$;
 \item for each scheduler $F$ there exists a scheduler $F'$ such that,
       for any classical channel $c \in cChan$
       $\con{Q || R}{\sigma} \Downarrow^p_F c$ implies that
       $\con{P || R}{\rho} \Downarrow^p_{F'} c$.
\end{enumerate}
\end{defi}

We can prove that $\approx_{oe}$ is an equivalence relation easily.

For example, we show two configurations that are not equivalent in the notion of 
open bisimulation but observationally equivalent.

\begin{exam}\label{examA}
Consider these two configurations:
\[
C = \con{M[q;x].(c!0 + d!0)}
{[\ket{+}]_q \otimes \rho},
\]
\[
D = \con{M[q;x].c!0 + M[q;x].d!0 + M[q;x].A(x)}
{[\ket{+}]_q \otimes \rho}
\]
where $A(x) := (\ifthen{x=0}{c!0})+(\ifthen{x=1}{d!0})$ and 
$M$ is as defined in Example \ref{examC}. 
The pLTSs for these configurations are depicted as in Figure \ref{fig:plts-exampleA}.
It is obvious that $C \not\approx_o D$.
However, we can prove that $C \approx_{oe} D$.

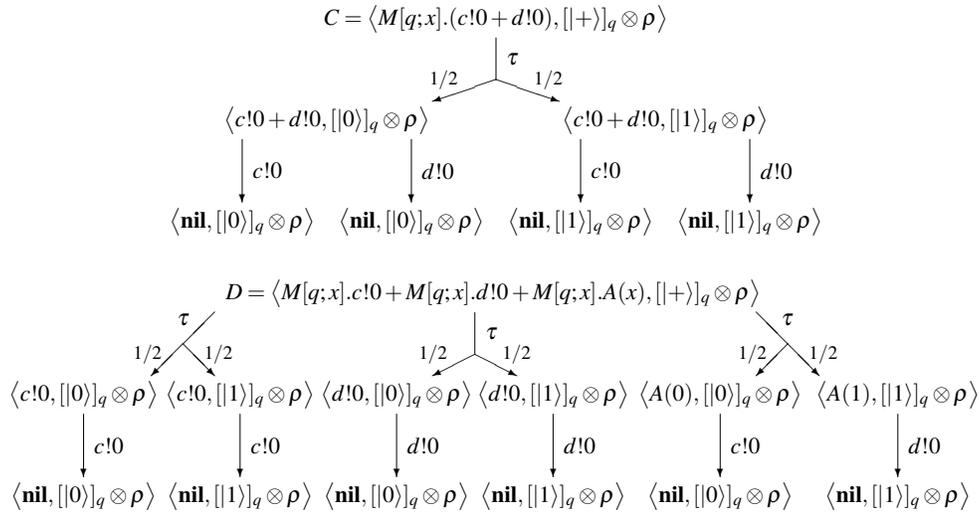
\begin{figure}[tb]
 \centering
 \scalebox{0.8}{\begin{picture}(460,238)(0,0)
  \put(75,130){
   \put(0,96){\makebox(310,12){%
    $C = \con{M[q;x].(c!0+d!0)}{[\ket{+}]_q \otimes \rho}$%
   }}
   \put(155,93){\line(0,-1){20}}
   \put(160,73){\makebox(30,20)[l]{$\tau$}}
   \put(155,73){\vector(-3,-1){30}}
   \put(107,68){\makebox(30,10)[r]{\footnotesize $1/2$}}
   \put(155,73){\vector(3,-1){30}}
   \put(173,68){\makebox(30,10)[l]{\footnotesize $1/2$}}
   \put(0,48){%
    \makebox(150,12){$\con{c!0+d!0}{[\ket{0}]_q \otimes \rho}$}%
   }
   \put(160,48){%
    \makebox(150,12){$\con{c!0+d!0}{[\ket{1}]_q \otimes \rho}$}%
   }
   \put(35,45){\vector(0,-1){30}}
   \put(40,15){\makebox(30,30)[l]{$c!0$}}
   \put(115,45){\vector(0,-1){30}}
   \put(120,15){\makebox(30,30)[l]{$d!0$}}
   \put(195,45){\vector(0,-1){30}}
   \put(200,15){\makebox(30,30)[l]{$c!0$}}
   \put(275,45){\vector(0,-1){30}}
   \put(280,15){\makebox(30,30)[l]{$d!0$}}
   \put(0,0){%
    \makebox(70,12){$\con{\mathbf{nil}}{[\ket{0}]_q \otimes \rho}$}%
   }
   \put(80,0){%
    \makebox(70,12){$\con{\mathbf{nil}}{[\ket{0}]_q \otimes \rho}$}%
   }
   \put(160,0){%
    \makebox(70,12){$\con{\mathbf{nil}}{[\ket{1}]_q \otimes \rho}$}%
   }
   \put(240,0){%
    \makebox(70,12){$\con{\mathbf{nil}}{[\ket{1}]_q \otimes \rho}$}%
   }
  }
  \put(0,0){
   \put(0,96){\makebox(460,12){%
    $D = \con{M[q;x].c!0 + M[q;x].d!0 + M[q;x].A(x)}{[\ket{+}]_q \otimes \rho}$
   }}
   \put(97,93){\line(-1,-1){15}}
   \put(55,82){\makebox(30,15)[r]{$\tau$}}
   \put(82,78){\vector(-1,-1){15}}
   \put(42,68){\makebox(30,10)[r]{\footnotesize $1/2$}}
   \put(82,78){\vector(1,-1){15}}
   \put(92,68){\makebox(30,10)[l]{\footnotesize $1/2$}}
   \put(220,93){\line(0,-1){20}}
   \put(225,73){\makebox(30,20)[l]{$\tau$}}
   \put(220,73){\vector(-2,-1){20}}
   \put(177,68){\makebox(30,10)[r]{\footnotesize $1/2$}}
   \put(220,73){\vector(2,-1){20}}
   \put(233,68){\makebox(30,10)[l]{\footnotesize $1/2$}}
   \put(353,93){\line(1,-1){15}}
   \put(365,82){\makebox(30,15)[l]{$\tau$}}
   \put(368,78){\vector(-1,-1){15}}
   \put(328,68){\makebox(30,10)[r]{\footnotesize $1/2$}}
   \put(368,78){\vector(1,-1){15}}
   \put(378,68){\makebox(30,10)[l]{\footnotesize $1/2$}}
   \put(0,48){%
    \makebox(70,12){$\con{c!0}{[\ket{0}]_q \otimes \rho}$}
   }
   \put(74,48){%
    \makebox(70,12){$\con{c!0}{[\ket{1}]_q \otimes \rho}$}
   }
   \put(148,48){%
    \makebox(70,12){$\con{d!0}{[\ket{0}]_q \otimes \rho}$}
   }
   \put(222,48){%
    \makebox(70,12){$\con{d!0}{[\ket{1}]_q \otimes \rho}$}
   }
   \put(296,48){%
    \makebox(80,12){$\con{A(0)}{[\ket{0}]_q \otimes \rho}$}
   }
   \put(380,48){%
    \makebox(80,12){$\con{A(1)}{[\ket{1}]_q \otimes \rho}$}
   }
   \put(35,45){\vector(0,-1){30}}
   \put(40,15){\makebox(30,30)[l]{$c!0$}}
   \put(109,45){\vector(0,-1){30}}
   \put(114,15){\makebox(30,30)[l]{$c!0$}}
   \put(183,45){\vector(0,-1){30}}
   \put(188,15){\makebox(30,30)[l]{$d!0$}}
   \put(257,45){\vector(0,-1){30}}
   \put(262,15){\makebox(30,30)[l]{$d!0$}}
   \put(336,45){\vector(0,-1){30}}
   \put(341,15){\makebox(30,30)[l]{$c!0$}}
   \put(420,45){\vector(0,-1){30}}
   \put(425,15){\makebox(30,30)[l]{$d!0$}}
   \put(0,0){%
    \makebox(70,12){$\con{\mathbf{nil}}{[\ket{0}]_q \otimes \rho}$}
   }
   \put(74,0){%
    \makebox(70,12){$\con{\mathbf{nil}}{[\ket{1}]_q \otimes \rho}$}
   }
   \put(148,0){%
    \makebox(70,12){$\con{\mathbf{nil}}{[\ket{0}]_q \otimes \rho}$}
   }
   \put(222,0){%
    \makebox(70,12){$\con{\mathbf{nil}}{[\ket{1}]_q \otimes \rho}$}
   }
   \put(296,0){%
    \makebox(80,12){$\con{\mathbf{nil}}{[\ket{0}]_q \otimes \rho}$}
   }
   \put(380,0){%
    \makebox(80,12){$\con{\mathbf{nil}}{[\ket{1}]_q \otimes \rho}$}
   }
  }
 \end{picture}}
 \caption{pLTSs for Example \ref{examA}}
 \label{fig:plts-exampleA}
\end{figure}

\end{exam}

\begin{prop}\label{propA}
Let $C$, $D$ be the configurations in Example \ref{examA}. Then 
$C \approx_{oe} D$.
\end{prop}

{\small
\begin{proof}
Let $P = M[q;x].(c!0 + d!0)$ and 
$Q = M[q;x].c!0 + M[q;x].d!0 + M[q;x].A(x)$. 

We have $qv(P) = qv(Q) = \{q\}$ and 
$\mathrm{tr}_{qv(P)}([\ket{+}]_q \otimes \rho) =
 \mathrm{tr}_{qv(Q)}([\ket{+}]_q \otimes \rho) = \rho$.

Let $R$ be an arbitrary quantum process. 
First, we need to show that, for each scheduler $F$, there exists a
scheduler $F'$ such that, for any classical channel $c$
$\con{P||R}{[\ket{+}]_q \otimes \rho} \Downarrow_F^p c$ implies 
$\con{Q||R}{[\ket{+}]_q \otimes \rho} \Downarrow_{F'}^p c$.
To prove it, we divide several cases of the scheduler $F$ and construct
a scheduler $F'$ in each case. 

\begin{enumerate}
\item
The scheduler $F$ does not choose the $\tau$ transition caused by $P$. 
In this case, we can easily construct a scheduler $F'$ such that 
$\con{Q||R}{[\ket{+}]_q \otimes \rho} \Downarrow_{F'}^p c$.
\item
The scheduler $F$ chooses the $\tau$ transition caused by $P$. 
In this case, we have
\[
\con{P||R}{[\ket{+}]_q \otimes \rho} \Longrightarrow_F 
\boxplus_{i \in I} \left(
\frac{1}{2}p_i \bullet \con{c!0+d!0||R_i}{[\ket{0}]_q \otimes \rho_i} \boxplus
\frac{1}{2}p_i \bullet \con{c!0+d!0||R_i}{[\ket{1}]_q \otimes \rho_i}
\right)
\]
after all $\tau$ transitions caused by $P$ and $R$ independently in accordance with $F$. 
Then, there exists a scheduler $F'$ such that 
\[
\con{Q||R}{[\ket{+}]_q \otimes \rho} \Longrightarrow_{F'} 
\boxplus_{i \in I} p_i \bullet \con{Q||R_i}{[\ket{+}]_q \otimes \rho_i}.
\]
For each $i \in I$, we again divide some cases of $F$ and construct $F'$ in each cases. 
Here we show only one case and omit the others. 

When 
\[
\con{c!0+d!0||R_i}{[\ket{0}]_q \otimes \rho_i} \trans{c!0}_F 
\con{\mathbf{nil}||R_i}{[\ket{0}]_q \otimes \rho_i},
\] 
\[
\con{c!0+d!0||R_i}{[\ket{1}]_q \otimes \rho_i} \trans{c!0}_F 
\con{\mathbf{nil}||R_i}{[\ket{1}]_q \otimes \rho_i},
\]
the channel $c$ is used with the probability $1$.
So, we can construct a scheduler $F'$ such that 
\[
\con{Q||R_i}{[\ket{+}]_q \otimes \rho_i} \trans{\tau}_{F'} \\
\frac{1}{2}\bullet\con{c!0||R_i}{[\ket{0}]_q \otimes \rho_i} \boxplus 
\frac{1}{2}\bullet\con{c!0||R_i}{[\ket{1}]_q \otimes \rho_i},
\]
\[
\con{c!0||R_i}{[\ket{0}]_q \otimes \rho_i} \trans{c!0}_{F'} 
\con{\mathbf{nil}||R_i}{[\ket{0}]_q \otimes \rho_i},
\hspace{3mm}
\con{c!0||R_i}{[\ket{1}]_q \otimes \rho_i} \trans{c!0}_{F'} 
\con{\mathbf{nil}||R_i}{[\ket{1}]_q \otimes \rho_i}.
\]
The scheduler $F'$ satisfies the requirement.
\end{enumerate}
\end{proof}
}

We show another example that means there exist configurations 
$C, D$ that $C \not\approx_{oe} D$ but 
$C \approx_o D$. 

\begin{exam}\label{examB}
Consider these two configurations:
\[
C = \con{M[q;x].A(q;x)}{[\ket{+}]_q \otimes \rho}, \hspace{5mm}
D = \con{c!0.\mathcal{I}[q]+d!0.\mathcal{I}[q]}
{[\ket{0}]_q \otimes \rho}
\]
where 
\[
A(q;x) 
:= \left(\ifthen{x=0}{(c!0.\mathcal{I}[q] + d!0.\mathcal{I}[q])}\right) + 
\left(\ifthen{x=1}{(c!0.\mathcal{X}[q] + d!0.\mathcal{X}[q])}\right),
\]
$\mathcal{I}$ is an operator that does nothing, 
$\mathcal{X}$ is the Pauli-$X$ operator, and 
$M$ is as defined in Example \ref{examC}.
The pLTSs for these configurations are depicted as in Figure \ref{fig:plts-exampleB}.

\begin{figure}[tb]
 \centering
 \scalebox{0.8}{\begin{picture}(350,291)(0,0)
  \put(0,130){%
   \put(0,149){\makebox(350,12){%
    $C = \con{M[q;x].A(q;x)}{[\ket{+}]_q \otimes \rho}$%
   }}
   \put(175,146){\line(0,-1){20}}
   \put(180,126){\makebox(30,20)[l]{$\tau$}}
   \put(175,126){\vector(-3,-1){40}}
   \put(125,118){\makebox(30,10)[r]{\footnotesize $1/2$}}
   \put(175,126){\vector(3,-1){40}}
   \put(193,118){\makebox(30,10)[l]{\footnotesize $1/2$}}
   \put(0,96){\makebox(170,12){%
    $\con{A(q;0)}{[\ket{0}]_q \otimes \rho}$%
   }}
   \put(180,96){\makebox(170,12){%
    $\con{A(q;1)}{[\ket{1}]_q \otimes \rho}$%
   }}
   \put(70,93){\vector(-1,-1){30}}
   \put(15,63){\makebox(30,30)[r]{$c!0$}}
   \put(100,93){\vector(1,-1){30}}
   \put(125,63){\makebox(30,30)[l]{$d!0$}}
   \put(250,93){\vector(-1,-1){30}}
   \put(195,63){\makebox(30,30)[r]{$c!0$}}
   \put(280,93){\vector(1,-1){30}}
   \put(305,63){\makebox(30,30)[l]{$d!0$}}
   \put(0,48){\makebox(80,12){%
    $\con{\mathcal{I}[q]}{[\ket{0}]_q \otimes \rho}$%
   }}
   \put(90,48){\makebox(80,12){%
    $\con{\mathcal{I}[q]}{[\ket{0}]_q \otimes \rho}$%
   }}
   \put(180,48){\makebox(80,12){%
    $\con{\mathcal{X}[q]}{[\ket{1}]_q \otimes \rho}$%
   }}
   \put(270,48){\makebox(80,12){%
    $\con{\mathcal{X}[q]}{[\ket{1}]_q \otimes \rho}$%
   }}
   \put(40,45){\vector(0,-1){30}}
   \put(45,15){\makebox(30,30)[l]{$\tau$}}
   \put(130,45){\vector(0,-1){30}}
   \put(135,15){\makebox(30,30)[l]{$\tau$}}
   \put(220,45){\vector(0,-1){30}}
   \put(225,15){\makebox(30,30)[l]{$\tau$}}
   \put(310,45){\vector(0,-1){30}}
   \put(315,15){\makebox(30,30)[l]{$\tau$}}
   \put(0,0){\makebox(80,12){%
    $\con{\mathbf{nil}}{[\ket{0}]_q \otimes \rho}$%
   }}
   \put(90,0){\makebox(80,12){%
    $\con{\mathbf{nil}}{[\ket{0}]_q \otimes \rho}$%
   }}
   \put(180,0){\makebox(80,12){%
    $\con{\mathbf{nil}}{[\ket{0}]_q \otimes \rho}$%
   }}
   \put(270,0){\makebox(80,12){%
    $\con{\mathbf{nil}}{[\ket{0}]_q \otimes \rho}$%
   }}
  }
  \put(100,0){
   \put(0,96){\makebox(150,12){%
    $D = \con{c!0.\mathcal{I}[q] + d!0.\mathcal{I}[q]}
     {[\ket{0}]_q \otimes \rho}$
   }}
   \put(35,93){\vector(0,-1){30}}
   \put(40,63){\makebox(30,30)[l]{$c!0$}}
   \put(115,93){\vector(0,-1){30}}
   \put(120,63){\makebox(30,30)[l]{$d!0$}}
   \put(0,48){\makebox(70,12){%
    $\con{\mathcal{I}[q]}{[\ket{0}]_q \otimes \rho}$%
   }}
   \put(80,48){\makebox(70,12){%
    $\con{\mathcal{I}[q]}{[\ket{0}]_q \otimes \rho}$%
   }}
   \put(35,45){\vector(0,-1){30}}
   \put(40,15){\makebox(30,30)[l]{$\tau$}}
   \put(115,45){\vector(0,-1){30}}
   \put(120,15){\makebox(30,30)[l]{$\tau$}}
   \put(0,0){\makebox(70,12){%
    $\con{\mathbf{nil}}{[\ket{0}]_q \otimes \rho}$%
   }}
   \put(80,0){\makebox(70,12){%
    $\con{\mathbf{nil}}{[\ket{0}]_q \otimes \rho}$%
   }}
  }
 \end{picture}}
 \caption{pLTSs for Example \ref{examB}}
 \label{fig:plts-exampleB}
\end{figure}
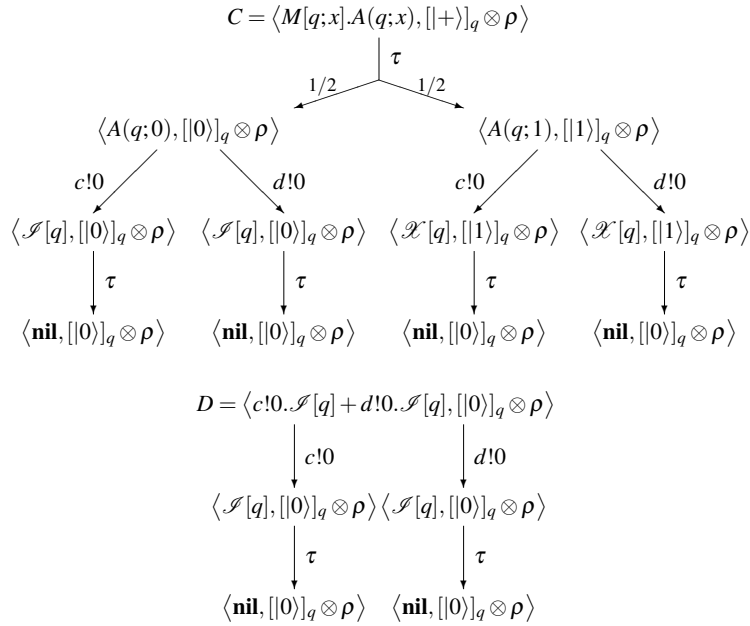

We can prove that $C \approx_o D$. 
However, $C \not\approx_{oe} D$. 
Consider a scheduler $F$ such that 
\[
F(\con{A(q;0)}{[\ket{0}]_q \otimes \rho}) = 
(c!0, \con{\mathcal{I}[q]}{[\ket{0}]_q \otimes \rho}),
\]
\[
F(\con{A(q;1)}{[\ket{1}]_q \otimes \rho}) = 
(d!0, \con{\mathcal{X}[q]}{[\ket{1}]_q \otimes \rho}).
\]
Then both $C\Downarrow_F^{1/2}c$ and $C\Downarrow_F^{1/2}d$ hold. 
But, for any schedulers $F'$, neither $D\Downarrow_{F'}^{1/2}c$ 
nor $D\Downarrow_{F'}^{1/2}d$ holds.
\end{exam}

\begin{prop}
$\approx_o$ and $\approx_{oe}$ are incomparable.
\end{prop}

\subsection{Strategy: a limited scheduler}

In previous section, we define schedulers and the observational equivalence. 
However, the processes in Example \ref{examC} are not observationally equivalent.
Consider a scheduler $F$ such that 
\[
F(\con{(c!0+d!0)}{[\ket{0}]_q \otimes \rho}) =
(c!0, \con{\mathbf{nil}}{[\ket{0}]_q \otimes \rho}),
\]
\[
F(\con{(c!0+d!0)}{[\ket{1}]_q \otimes \rho}) =
(d!0, \con{\mathbf{nil}}{[\ket{1}]_q \otimes \rho}).
\]
Then both $C\Downarrow_F^{1/2}c$ and $C\Downarrow_F^{1/2}d$ hold. 
But, for any schedulers $F'$, neither $D\Downarrow_{F'}^{1/2}c$ 
nor $D\Downarrow_{F'}^{1/2}d$ holds.

This problem is due to the definition of schedulers, that is, because schedulers can 
choose different transitions even though the processes are the same.
In order to solve this problem, we propose strategies, limited schedulers.

\begin{defi}\label{def:strategy}
A function $F: Con \to (Act \times D(Con)) \cup \{\bot\}$ is a \textit{strategy} 
if the following conditions are satisfied:
\begin{enumerate}
\item
$F(C) = (\alpha, \mu)$ implies $C \trans{\alpha} \mu$,
\item
$F(C) = \bot$ implies $C \not\rightarrow$,
\item
if $F(\con{P}{\rho}) = (\alpha, \mu)$, then there exist a set of processes 
$\{P_i\}_{i \in I}$, a set of super-operators 
$\{\mathcal{E}_i\}_{i \in I}$, acting on $\mathcal{H}_{qv(P)}$, and a set 
of projectors $\{E_i\}_{i \in I}$, acting on $\mathcal{H}_{qv(P)}$ and 
$\sum_{i \in I} E_i = I$, such that for any density operators $\sigma$,
\[
F(\con{P}{\sigma}) = 
\left(\alpha, 
      \sum_{i \in I \wedge q_i^\sigma \not= 0}
      q_i^\sigma \con{P_i}{\mathcal{E}_i(\sigma)/q_i^\sigma} \right)
\]
and
\[
\mu = \sum_{i \in I \wedge q_i^\rho \not= 0}
      q_i^\rho \con{P_i}{\mathcal{E}_i(\rho)/q_i^\rho}
\]
where $q_i^\sigma = \mathrm{tr}(E_i\sigma)$.
\end{enumerate}
\end{defi}

The difference between schedulers and strategies is only the condition 3 in 
Definition \ref{def:strategy}. This condition means that strategies must choose the same 
transition for any density operators if the processes of the configurations are the same. 
In order to validate this condition, we use the following lemma. This lemma is 
stronger than Lemma 3.3 (2) in \cite{Feng2012}, but can
still be easily observed from the transition rules of qCCS.

\begin{lemm}
If $\con{P}{\rho} \trans{\alpha} \mu$, then there exists a set of processes 
$\{P_i\}_{i \in I}$, a set of super-operators 
$\{\mathcal{E}_i\}_{i \in I}$, acting on $\mathcal{H}_{qv(P)}$, and a set 
of projectors $\{E_i\}_{i \in I}$, acting on $\mathcal{H}_{qv(P)}$ and 
$\sum_{i \in I}E_i = I$, such that for any density operators $\sigma$,
\[
\con{P}{\sigma} \trans{\alpha} 
\sum_{i \in I \wedge q_i^\sigma \not= 0}
q_i^\sigma \con{P_i}{\mathcal{E}_i(\sigma)/q_i^\sigma},
\]
and
\[
\mu = \sum_{i \in I \wedge q_i^\rho \not= 0}
q_i^\rho \con{P_i}{\mathcal{E}_i(\rho)/q_i^\rho}
\]
where $q_i^\sigma = \mathrm{tr}(E_i\sigma)$.
\end{lemm}

We use the notations $C \trans{\alpha}_F \mu$, $C \trans{\alpha}_F$ 
and $\Longrightarrow_F$ for strategies $F$ in the same way as schedulers. 

\subsection{Observational equivalence with strategies}

We write $C \Downarrow_F^p c$ for strategies $F$ in the same way as schedulers. 
Now, we define observational equivalence using strategies instead of schedulers. 

\begin{defi}
Two configurations $\con{P}{\rho}, \con{Q}{\sigma} \in Con$ are 
\textit{observationally equivalent with strategies}, 
we write $\con{P}{\rho} \approx_{oe}^{st} \con{Q}{\sigma}$, 
if $qv(P) = qv(Q)$, $\mathrm{tr}_{qv(P)}(\rho) = \mathrm{tr}_{qv(Q)}(\sigma)$ and 
for any quantum processes $R \in qProc$,
\begin{enumerate}
 \item for each strategy $F$ there exists a strategy $F'$ such that,
       for any classical channel $c \in cChan$
       $\con{P || R}{\rho} \Downarrow^p_F c$ implies that
       $\con{Q || R}{\sigma} \Downarrow^p_{F'} c$;
 \item for each strategy $F$ there exists a strategy $F'$ such that,
       for any classical channel $c \in cChan$
       $\con{Q || R}{\sigma} \Downarrow^p_F c$ implies that
       $\con{P || R}{\rho} \Downarrow^p_{F'} c$.
\end{enumerate}
\end{defi}
We can prove that $\approx_{oe}^{st}$ is an equivalence relation easily.

Now, we can check that the two configurations in Example \ref{examC} are observationally 
equivalent with strategies.

\begin{prop}\label{propC}
Let $C$ and $D$ be configurations in Example \ref{examC}.
Then $C \approx_{oe}^{st} D$.
\end{prop}

Let us consider the relation among open bisimulation $\approx_o$, 
observational equivalence $\approx_{oe}$, and observational equivalence with strategies 
$\approx_{oe}^{st}$.

By Example \ref{examC} and Proposition \ref{propC}, 
there exist some configurations $C$ and $D$ that 
$C \not\approx_{oe} D$ but 
$C \approx_{oe}^{st} D$. 
However, $\approx_{oe} \subset \approx_{oe}^{st}$ does not hold. 
Consider Example \ref{examA} again.
The configurations in Example \ref{examA} are observationally equivalent, but 
they are not observationally equivalent with strategies. 
Consider the strategy $F$ such that 
\[
F(D) = 
\left( 
\tau , 
\frac{1}{2}\bullet\con{A(0)}{[\ket{0}]_q\otimes\rho} \boxplus
\frac{1}{2}\bullet\con{A(1)}{[\ket{1}]_q\otimes\rho}
\right),
\]
\[
F(\con{A(0)}{[\ket{0}]_q\otimes\rho}) = 
\left(
c!0 , 
\con{\mathbf{nil}}{[\ket{0}]_q\otimes\rho}
\right),
\hspace{5mm}
F(\con{A(1)}{[\ket{1}]_q\otimes\rho}) = 
\left(
d!0, 
\con{\mathbf{nil}}{[\ket{1}]_q\otimes\rho}
\right).
\]
Then both $D\Downarrow_F^{1/2}c$ and $D\Downarrow_F^{1/2}d$ are hold. 
However, neither $C\Downarrow_{F'}^{1/2}c$ nor 
$C\Downarrow_{F'}^{1/2}d$ holds for any strategies $F'$.
It is because, for any strategies $F'$, if 
\[
F'(\con{c!0+d!0}{[\ket{0}]_q\otimes\rho}) = (\alpha_0, \mu_0),
\hspace{5mm}
F'(\con{c!0+d!0}{[\ket{1}]_q\otimes\rho}) = (\alpha_1, \mu_1),
\]
then $\alpha_0$ and $\alpha_1$ must be the same action by 
the definition of strategies. 

\begin{prop}
$\approx_{oe}$ and $\approx_{oe}^{st}$ are incomparable.
\end{prop}

In addition, $\approx_o \subset \approx_{oe}^{st}$ does not also hold, although there 
exist some configurations $C$ and $D$ that 
$C \not\approx_o D$ but 
$C \approx_{oe}^{st} D$.
Consider Example \ref{examB} again. It is proved that $C \approx_o D$, 
but $C \not\approx_{oe}^{st} D$. Consider the strategy $F$ such that 
\[
F(C) = \left( \tau, 
\frac{1}{2}\bullet\con{A(q;0)}{[\ket{0}]_q\otimes\rho} \boxplus 
\frac{1}{2}\bullet\con{A(q;1)}{[\ket{1}]_q\otimes\rho}
\right),
\]
\[
F(\con{A(q;0)}{[\ket{0}]_q\otimes\rho}) = \left( c!0, 
\con{\mathcal{I}[q]}{[\ket{0}]_q\otimes\rho}
\right),
\]
\[
F(\con{A(q;1)}{[\ket{1}]_q\otimes\rho}) = \left( d!0,
\con{\mathcal{X}[q]}{[\ket{1}]_q\otimes\rho}
\right).
\]
Then both $C\Downarrow_F^{1/2}c$ and $C\Downarrow_F^{1/2}d$ are hold. 
However, neither $D\Downarrow_{F'}^{1/2}c$ nor 
$D\Downarrow_{F'}^{1/2}d$ holds for any strategies $F'$. 

\begin{prop}
$\approx_o$ and $\approx_{oe}^{st}$ are incomparable.
\end{prop}

\section{Related work}

There already exists ``observational equivalence'' or ``observational
congruence'' on other process calculi such as applied pi calculus
\cite{Abadi2001} and probabilistic applied pi calculus
\cite{Goubault-larrecq2007}. However, they are essentially the same as
reduction barbed congruence because they are reduction-closed by
definition. So, they are also the same as the notion of open
bisimulation.

The same notion of observational equivalence in this paper is defined along
the line of applied pi calculi in \cite{Kubota2011}. However, the study in
\cite{Kubota2011} was not so sophisticated and many unsolved problems
were taken over by this study.

On the other hand, probabilistic branching bisimilarity, another notion
of behavioral equivalence, is defined on CQP in
\cite{Davidson2012a, Davidson2012}. The same idea as strategies in our
work is used in its definition.

\section{Conclusion}

In this paper, we proposed the notion of observational equivalence. 
To define it, we used schedulers that solve nondeterministic choices. 
Some processes that are not bisimilar became observationally equivalent,
but others remained nonequivalent. And so, we defined strategies, which
are limited schedulers, and the notion of observational equivalence with
strategies. Some processes that are intuitively equivalent became
observationally equivalent with strategies. After that, we investigated
the relation among three notions, that is, open bisimulation
$\approx_o$, observational equivalence $\approx_{oe}$, and observational
equivalence with strategies $\approx_{oe}^{st}$, 
and we found that it is impossible to compare these three notions. 
Even so, we think that $\approx_{oe}^{st}$ is the most intuitive of the
three when we consider the situation like Example \ref{examC} or the
formal security proof of BB84.

However, there remains a question whether our definition of
observational equivalence is really intuitive. In order to solve this
question, we must formalize the ``intuition'' at first. And then, we can
discuss whether our definition of equivalence is intuitive or not.

We should also discuss the congruence of our observational
equivalences. Congruence is the property that the equivalence is
preserved under process constructs. The congruence property for parallel
compositions $P || R$, which are the most important case, holds by
definition of our observational equivalences. In addition, the property
for relabelling functions $P[f]$ and conditional executions
$\ifthen{b}{P}$ also holds. However, the property for channel
restrictions $P \backslash L$ does not hold. For example,
$\con{c!0 + d!0}{\rho} \approx_{oe} \con{\tau. c!0 + \tau. d!0}{\rho}$
but
$\con{(c!0 + d!0)\backslash\{c\}}{\rho} \not\approx_{oe}
\con{(\tau. c!0 + \tau. d!0) \backslash \{c\}}{\rho}$
for any density operator $\rho$.
It remains for future work to investigate whether they are preserved
under other constructs or not.


\bibliographystyle{eptcs}
\bibliography{qccs-observational-equiv_qpl2014_post}

\end{document}